\documentclass{ws-procs9x6-cpt16}
\begin{document}

\newcommand{\refeq}[1]{(\ref{#1})}
\def\etal {{\it et al.}}

\title{Constraints on SME Coefficients from Lunar Laser Ranging,\\ 
Very Long Baseline Interferometry, and \\
Asteroid Orbital Dynamics}

\author{C.\ Le Poncin-Lafitte,$^1$ 
A.\ Bourgoin,$^1$ A.\ Hees,$^2$ S.\ Bouquillon,$^1$\\
S.\ Lambert,$^1$ G.\ Francou,$^1$ M.-C.\ Angonin,$^1$ Q.G.\ Bailey,$^3$\\ 
D.\ Hestroffer,$^4$ P.\ David,$^4$ F.\ Meynadier,$^1$ and P.\ Wolf$^1$ }

\address{$^1$SYRTE, Observatoire de Paris, PSL Research University, CNRS\\
Sorbonne Universit\'es, UPMC Univ.\ Paris 06, LNE\\
61 avenue de l'Observatoire, 75014 Paris, France}

\address{$^2$Department of Physics and Astronomy, UCLA, 
Los Angeles, CA 90095, USA}

\address{$^3$Department of Physics, Embry-Riddle Aeronautical University\\
3700 Willow Creek Road, Prescott, AZ 86301, USA}

\address{$^4$ IMCCE, Observatoire de Paris, PSL Research University, CNRS  \\
Sorbonne Universit\'es, UPMC Univ.\ Paris 06, Univ.\ de Lille\\
77 avenue Denfert-Rochereau, 75014 Paris, France}

\begin{abstract}
Lorentz symmetry violations can be parametrized by an effective field theory framework that contains both General Relativity and the Standard Model of particle physics, called the Standard-Model Extension or SME. We consider in this work only the pure gravitational sector of the minimal SME. We present new constraints on the SME coefficients obtained from lunar laser ranging, very long baseline interferometry, and planetary motions.
\end{abstract}

\bodymatter

\section{Introduction}

The solar system remains the most precise laboratory to test the theory of gravity, that is to say General Relativity (GR). Constraints on deviations from GR can only be obtained in an extended theoretical framework that parametrizes such deviations.  The parametrized post-Newtonian formalism is one of them and has been widely used for decades. More recently, other phenomenological frameworks have been developed like the Standard-Model Extension (SME), which is an extensive formalism that allows a systematic description of Lorentz symmetry violations in all sectors of physics, including gravity. We present here new constraints on pure-gravity sector coefficients of the minimal SME obtained with very long baseline interferometry (VLBI), lunar laser ranging (LLR) and planetary motions. We also assess the possibility to constrain them using future asteroids observations by Gaia.

\section{Very long baseline interferometry}

VLBI is a geometric technique which measures the time difference in the arrival of a radio wavefront emitted by a distant radio source (typically a quasar) between at least two Earth-based radio telescopes, with a precision of a few picoseconds. Although this technique is initially dedicated to tracking the Earth's rotation and enabling the realization of global reference frames, it allows also performing fundamental physics tests by measuring the relativistic bending of light rays due to the Sun and the planets.\cite{2009A&A...499..331L,2011A&A...529A..70L} Recently, the VLBI gravitational group delay has been derived in the SME formalism (see Eqs.~(7) and~(10) from Ref.\ \refcite{2016arXiv160401663L}). Using observations between August 1979 and mid-2015 consisting of almost 6000 VLBI 24-hr sessions (corresponding to 10 million delays), we turned to a global solution in which we estimated $\bar s^{TT}$ as a global parameter together with radio source coordinates. We obtained\cite{2016arXiv160401663L}
\begin{equation}
    \bar s^{TT}=(-5\pm8)\times10^{-5}\, ,
\end{equation} with a global postfit rms of 28~ps and a $\chi^2$ per degree of freedom of 1.15. Correlations between radio source coordinates and $\bar s^{TT}$ are lower than 0.02, the global estimate being consistent with the mean value obtained with the session-wise solution with a slightly lower error.

\section{Lunar laser ranging}
Some years ago, a first estimate of SME coefficients with LLR data has been obtained.\cite{2007PhRvL..99x1103B} However it was a fit of theoretical SME signatures in residuals of LLR measurements analyzed previously in pure GR. This kind of approach is not fully satisfactory and provides order of magnitude upper limits on SME coefficients but not real estimates or constraints on them (see the discussion in Ref.\ \refcite{2016arXiv160401663L}). Therefore, we built a new numerical lunar ephemeris called the {\it Eph\'em\'eride Lunaire Parisienne Num\'erique} (ELPN) computed in the SME framework, taking into account effects of the SME on the orbital dynamics and on the propagation of light.\cite{2006PhRvD..74d5001B} A global adjustment to LLR observations allows us to estimate properly some linear combinations of SME coefficients, as illustrated in Table~\ref{tab1}  (see also Ref.\ \refcite{2016arXiv160700294B}). Those constraints take into account correlations between SME coefficients and other parameters (such as positions, velocities, masses, \dots) and are more reliable than the first analysis from Ref.\ \refcite{2007PhRvL..99x1103B}. 

\begin{table} 
\tbl{Estimated values of SME coefficients obtained from LLR.\cite{2016arXiv160700294B}}
{  \begin{tabular}{ccc}
\toprule
	SME   coefficients           & LLR estimation from Ref.\ \refcite{2016arXiv160700294B} \\
	\hline            
	$\bar{s}^{T\!X}$ &$(-0.9\pm1.0)\times10^{-8\ }$ \\
	$\bar{s}^{X\!Y}$ &$(-5.7\pm7.7)\times10^{-12}$  \\
	$\bar{s}^{X\!Z}$ &$(-2.2\pm5.9)\times10^{-12}$  \\
	$\bar{s}^{XX}-\bar{s}^{YY}$      &$(+0.6\pm4.2)\times10^{-11}$  \\
	$\bar{s}^{TY}+0.43 \bar s^{TZ}$      &$(+6.2\pm7.9)\times10^{-9\ }$ \\
	$\bar{s}^{YZ}-22.2\left(\bar s^{XX}+\bar s^{YY}-2\bar s^{ZZ}\right)$      &$(-0.5\pm1.0)\times10^{-9\ }$ \\
	\botrule          
  \end{tabular}
 }
  \label{tab1}
\end{table}

\section{Gaia observations of solar system objects}

Launched in December 2013, the ESA Gaia mission is scanning the whole celestial sphere once every 6 months providing high precision astrometric data for a huge number ($\approx$ 1 billion) of celestial bodies. In addition to stars, it is also observing solar system objects. In particular, about 360,000 asteroids will regularly be observed at the sub-mas level. We simulated the trajectories of 10,000 asteroids within the SME framework and performed a realistic covariance analysis taking into account the Gaia trajectory and scanning law (see Ref.\ \refcite{mouret} for more details about the strategy). The covariance analysis leads to the estimated uncertainties presented in Table~\ref{tab:SME} (see Ref.\ \refcite{2015sf2a.conf..125H}). These uncertainties are better than the current best estimations of the SME parameters available in the literature.\cite{2011RvMP...83...11K} In particular, they are better than ones obtained with planetary ephemerides.\cite{2015PhRvD..92f4049H} This is due to the variety of asteroid orbital parameters while planetary ephemerides use only 8 planets with similar orbital parameters (same orbital planes and nearly circular orbits). Therefore, the estimation of the SME coefficients with planetary ephemerides are highly correlated, which degrades the marginalized SME estimates (see the discussion in Ref.\ \refcite{2015PhRvD..92f4049H}). Using our set of asteroids, the correlation matrix for the SME coefficients is very reasonable: the three most important correlation coefficients are 0.71, $-0.68$, and 0.46. All the other correlations are below 0.3. Therefore, Gaia offers a unique opportunity to constrain Lorentz violation through the SME formalism.
\begin{table}[hbt]
\tbl{Sensitivity on SME parameters.}
{
\begin{tabular}{c c}          
\toprule                        
SME coefficients & Sensitivity \\    
\hline                                   
     $\bar s^{XX}-\bar s^{YY}$  & $9\times 10^{-12}$ \\      
     $\bar s^{XX}+\bar s^{YY}-2\bar s^{ZZ}$ &  $2\times 10^{-11}$ \\
     $\bar s^{XY}$ & $4\times 10^{-12}$\\
     $\bar s^{XZ}$ & $2\times 10^{-12}$ \\
     $\bar s^{YZ}$ & $4\times 10^{-12}$ \\
     $\bar s^{TX}$ & $1\times 10^{-8\phantom{1}}$\\
     $\bar s^{TY}$ & $2\times 10^{-8\phantom{1}}$\\
     $\bar s^{TZ}$ & $4\times 10^{-8\phantom{1}}$ \\
\botrule                                             
\end{tabular}
}
\label{tab:SME}  
\end{table}

\section{Conclusion}

We presented our latest constraints on gravity-sector SME coefficients obtained with LLR and VLBI observations. We highlighted also the improvement that we can expect from Gaia observations of asteroids in the future. A combined analysis with planetary ephemerides analysis,\cite{2015PhRvD..92f4049H} Lunar Laser Ranging,\cite{2007PhRvL..99x1103B,2016arXiv160700294B} atom interferometry,\citep{2008PhRvL.100c1101M} and binary pulsars\cite{2014PhRvL.112k1103S} would also be very interesting in order to decorrelate almost all gravity-sector SME coefficients and produce the most stringent estimate on the SME coefficients. Our analysis needs to be extended to include gravity-matter Lorentz violation in the SME framework.

\section*{Acknowledgments}

Q.G.B.\ acknowledges financial support from the NSF Grant No.\ PHY-1402890 and from Sorbonne Universit\'es through an ``Emergence'' grant.

\end{document}